\documentclass[
reprint,
superscriptaddress,
amsmath,amssymb,
aps,pre
]{revtex4-2}

\usepackage{amsfonts}

\usepackage{graphicx}
\usepackage{dcolumn}
\usepackage{bm}
\usepackage{hyperref}
\usepackage{physics}
\usepackage{xcolor}
\usepackage{mathrsfs}
\usepackage{mathtools}
\usepackage{comment}
\begin{document}

\title{Universality in long-range interacting systems: the effective dimension approach}

\author{Andrea Solfanelli}
\email{asolfane@sissa.it}
\affiliation{SISSA, via Bonomea 265, 34136 Trieste, Italy}
\affiliation{INFN, Sezione di Trieste, via Valerio 2, 34127 Trieste, Italy}

\author{Nicol\`o Defenu}
\affiliation{Institut f\"ur Theoretische Physik, ETH Z\"urich, Wolfgang-Pauli-Str. 27 Z\"urich, Switzerland}

\date{\today} 
\begin{abstract}
Dimensional correspondences have a long history in critical phenomena. Here, we review the effective dimension approach, which relates the scaling exponents of a critical system in $d$ spatial dimensions with power-law decaying interactions $r^{d+\sigma}$ to a local system, i.e. with finite range interactions, in an effective fractal dimension $d_\mathrm{eff}$. This method simplifies the study of long-range models by leveraging known results from their local counterparts. While the validity of this approximation beyond the mean-field level has been long debated, we demonstrate that the effective dimension approach, while approximate for non-Gaussian fixed points, accurately estimates the critical exponents of long-range models with an accuracy typically larger than $97\%$. To do so, we review perturbative RG results, extend the approximation's validity using functional RG techniques, and compare our findings with precise numerical data from conformal bootstrap for the two-dimensional Ising model with long-range interactions.
\end{abstract}

\maketitle

\section{Introduction}
The universality of critical phenomena is one of the most relevant fields of investigation in modern physics. Thanks to the universal nature of continuous phase transitions, the same formalism can be applied to phase transitions that occur in vastly different physical systems, all belonging to the same class of universality. the general renormalization-group (RG) framework elucidates the critical behavior of many systems and their universal features. Since its inception\,\cite{Wilson1971renormalizationI,Wilson1971renormalizationII}, critical phenomena have been extensively studied, leading to the development of numerous ideas aimed at comprehending the critical behavior of increasingly complex systems, ranging from ferromagnets to the standard model of high-energy particle physics\,\cite{Zinn2021quantum,Pellissetto2002critical}.

For several decades, this understanding was largely confined to systems exhibiting local, finite-range interactions, such as lattice systems with nearest-neighbor couplings or local field theories. However, in more recent times, substantial efforts have been directed towards understanding the critical features of both classical and quantum long-range (LR) interacting models\,\cite{Campa2009Statistical,Campa2014physics,Defenu2023Longrange}. These models describe physical systems with couplings among pairs of microscopic constituents $V(r)$ that decay as a power law of their distance: $V(r)\sim 1/r^{d+\sigma}$ in the large $r$ limit ($r\to\infty$), where $d$ represents the spatial dimension of the system. Notably the presence of such non-local interactions may alter the standard picture provided by the Mermin-Wagner theorem\,\cite{Mermin1966absence} allowing for transition at dimensions smaller than the lower critical one for usual local models\,\cite{Fisher1972critical}. Moreover, when $\sigma$ is less than 0, the energy density diverges, requiring the rescaling of the coupling constant to properly define the thermodynamic limit\,\cite{Campa2009Statistical}. While, when $\sigma>0$, the model often displays a second order phase transition. Specifically, depending on the parameter $\sigma>0$, three distinct regimes can be identified\,\cite{Defenu2020criticality}: 
\begin{enumerate}
    \item For $\sigma\leq\sigma_{\mathrm{mf}}$, where $\sigma_{\mathrm{mf}}$ can be calculated in the mean-field approximation, the mean-field approximation correctly describes the universal behavior.
    \item For $\sigma_{\mathrm{mf}}< \sigma\leq \sigma^*$ the system exhibits peculiar LR critical exponents.
    \item For $\sigma > \sigma^*$ the critical behavior corresponds to the local case ($\sigma\to\infty$).
\end{enumerate}

An intriguing perspective for interpreting these findings is the so called effective dimension approach. This concept suggests the potential to deduce most critical properties of a LR model in dimension $d$ with a power-law exponent $d + \sigma$ from those of a local model in an effective fractional dimension $d_{\mathrm{eff}}$, linked to $d$ and $\sigma$ through the relation\,\cite{Kotliar1983one, Banos2012correspondence, Angelini2014relations}
\begin{align}
    d_{\mathrm{eff}} = \frac{d(2-\eta_{\mathrm{SR}}(d_{\mathrm{eff}}))}{\sigma},\label{eq: effective dim}
\end{align}
where $\eta_{\mathrm{SR}}(d_{\mathrm{eff}})$ is the $\eta$ critical exponent of the local model ($\sigma\to\infty$) in $d_{\mathrm{eff}}$ dimension. One of the primary advantages of this approach is its capability to reproduce the behavior within and beyond the validity range of mean-field approximation by varying a single parameter. This proves highly advantageous for numerical simulations, as the computational complexity of the model remains constant when we change the effective dimension \cite{Angelini2014relations}.
While the result in Eq.\,\eqref{eq: effective dim} holds true at the mean-field level and at leading order in perturbation theory\,\cite{Joyce1966spherical}, its approximate nature has been evidenced in several calculations\,\cite{Defenu2015fixedpoint,Behan2017scaling}. 

In this study, we aim to provide a step forward in resolving this longstanding debate. We demonstrate that the effective dimension expression in Eq.\,\eqref{eq: effective dim}, although not exact for non-Gaussian fixed points, serves as a highly accurate approximation. It offers an exceedingly efficient means to estimate critical exponents, exhibiting a minimal error, with an accuracy, which we estimate to be higher than $97\%$, even in proximity of $\sigma^*$, when compared to precise numerical estimates.

The paper is organized as follows. In Sec.\,\ref{sec: derivation of deff}, we briefly review how to obtain the effective dimension relation in Eq.\,\eqref{eq: effective dim} by means of scaling theory, demonstrating that it holds true at leading order in perturbative RG around the Gaussian fixed point. In Sec.\,\ref{sec: Perturbative RG one loop results} we apply know perturbative RG results in oder to obtain an explicit formula for $d_{\rm eff}$ valid up to corrections of order $\mathcal{O}(\varepsilon^3)$. In Sec.\,\ref{sec: Functional RG results}, we introduce a more modern approach to the problem based on functional RG techniques. This approach extends the validity of Eq.\,\eqref{eq: effective dim} beyond the one-loop approximation by incorporating corrections from wavefunction renormalization. Finally, in Sec.\,\ref{sec: Comparison with exact numerics}, we compare the predictions of the effective dimension approach with conformal bootstrap numerical data for the two-dimensional Ising model with LR interactions, demonstrating that the accuracy predicted by functional RG remains consistent at the exact level. Based on the latter result, we argue that, at least for $\mathrm{O}(N)$ vector models, the accuracy of the effective dimension approach will remain above $97\%$.

\section{Effective dimension and scaling theory}
\label{sec: derivation of deff}
We begin our analysis by showing how the effective dimension relation in Eq.\,\eqref{eq: effective dim} can be obtained by scaling theory. Our prototypical model is the family of classical $\mathrm{O}(N)$-symmetric spin models, whose Hamiltonian is given by
\begin{align}
\label{h_ON}
    H = -\frac{1}{2}\sum_{i\neq j}J_{i,j}\mathbf{S}_i\cdot\mathbf{S}_j,
\end{align}
where $\mathbf{S}_i$ is an $N$-component spin vector with unit modulus, $J_{ij} > 0$ are ferromagnetic translational invariant couplings, and the indices $i$ and $j$ run over all sites on any $d$-dimensional regular lattice of $V$ sites. Specifically, we consider the case of power law decaying couplings $J_{i,j} = J/r_{ij}^{d + \sigma}$, where $r$ is the distance between two sites of the system. From a field-theoretic perspective, the Hamiltonian in Eq.\,\eqref{h_ON} lies in the same universality class of the continuous action
\begin{align}
   S = \int \frac{d^dk}{(2\pi)^d}\omega(k)|\boldsymbol{\varphi}(k)|^2+u\int d^dx|\boldsymbol{\varphi}(x)|^4,\label{eq: action}
\end{align}
where the dispersion relation encodes only the low-energy momentum contribution $\omega(k) = a_\sigma k^\sigma + a_2 k^2 + r$ and $\boldsymbol{\varphi}$ is an $N$-component bosonic field\,\cite{Sak1973recursion}.
At a Gaussian level $(u = 0)$, LR interactions become relevant for $\sigma<2$, so that, $\sigma^* = 2$. The corresponding length dimension of the bosonic field is given by
\begin{align}
    [\varphi]\sim \begin{cases}
        L^{-(d-\sigma)/2} &\mathrm{for}\quad\sigma<2\\
        L^{-(d-2)/2} &\mathrm{for}\quad\sigma>2
    \end{cases}.
\end{align}
If follows that, at criticality, for $\sigma<2$, we have
\begin{align}
    \langle\varphi(x)\varphi(0)\rangle\sim \frac{1}{x^{d-\sigma}} = \frac{1}{x^{d-2+\eta_{\rm LR}}},
\end{align}
where $\eta_{\rm LR} = 2-\sigma$ is the anomalous dimension computed with respect to the canonical dimension of the local theory. It can be thought of as a measure of how the decay of correlations differs from the local one.

Let us now introduce interactions into our framework as a perturbation of the Gaussian theory. Considering the quartic term in the action \eqref{eq: action}, we have $[\varphi^4]\sim L^{-2(d-\sigma)}$ for $\sigma<2$, so that the perturbation is irrelevant as long as $\sigma < d/2$. Accordingly, in this regime, mean-field results are exact, and therefore, for $\mathrm{O}(N)$ models, $\sigma_{\mathrm{mf}} = d/2$. To probe values of $\sigma$ outside the mean-field region, one can use perturbative RG around the $d = 4, \sigma = 2$ Gaussian fixed point, expanding in terms of $\varepsilon = 2\sigma-d$. This problem was addressed in the seminal papers  \cite{Fisher1972critical,Sak1973recursion}. One of the main findings of these studies is the observation that the $\propto k^{\sigma}|\boldsymbol{\varphi}(k)|^2$ term in the action\,\eqref{eq: action} does not acquire an anomalous scaling. Intuitively, this may be traced back to the fact that the perturbative expansion can only generate integer powers of $k^2$, not affecting the non-analytic $\propto k^\sigma$ behavior.
As a consequence, even in the presence of interactions, the scaling dimension of the LR kinetic term in the Hamiltonian is given by $\Delta_\sigma = 2\Delta_{\varphi}+\sigma$. 

However, in presence of interactions the actual scaling of the local theory shall read $\Delta_{\varphi} = (d - 2 + \eta_{\mathrm{SR}})/2$, so that $\Delta_\sigma = d +\eta_{\mathrm{SR}}-\eta_{\mathrm{ LR}}$\,\cite{Sak1973recursion}. Thus, the boundary between LR and local behaviour has to be located at the value $\sigma^*$ such that $\eta_{\mathrm{ LR}}(\sigma)=\eta_{\mathrm{SR}}$.
Thus, we can compare the scaling behaviour of the two interacting theories in order to justify Eq.\,\eqref{eq: effective dim}. 

It is convenient to follow the procedure introduced in Refs.\,\cite{Banos2012correspondence,Angelini2014relations}and consider the general scaling form of the singular part of the free energy density for a LR system in $d$ dimensions and a local system in $d_{\mathrm{eff}}$ dimensions and equate them, leading to
\begin{align}
    f_s &= \frac{1}{V}\Phi_{\mathrm{LR}}(V^{y_{\tau}^{\mathrm{LR}}/d}\tau,V^{y_{h}^{\mathrm{LR}}/d}h,V^{y_{u}^{\mathrm{LR}}/d}u)\notag\\
    &=\frac{1}{V}\Phi_{\mathrm{SR}}(V^{y_{\tau}^{\mathrm{SR}}/d_{\mathrm{eff}}}\tau,L^{y_{h}^{\mathrm{SR}}/d_{\mathrm{eff}}}h,V^{y_{u}^{\mathrm{SR}}/d_{\mathrm{eff}}}u),
\end{align}
where $V$ is the total number of spins, $\tau$ is the reduced temperature, $h$ is the reduced magnetic field, and $u$ is the coupling of the irrelevant operator that gives the leading corrections. The exponents $y_\tau$, $y_h$, $y_u$ are connected to the eigenvalues of the linearized form of the RG transformation around the critical fixed point. Thus, the connection between the exponents is 
\begin{align}
    y^{\mathrm{LR}}/d = y^{\mathrm{SR}}/d_{\mathrm{eff}}.\label{eq: relation y}
\end{align}
Combining this condition with the relations of the 
$y$s with the critical exponents, we obtain \cite{Angelini2014relations}
\begin{align}
    &d\nu_{\mathrm{LR}} = d_{\mathrm{eff}}\nu_{\mathrm{SR}},\quad \frac{2-\eta_{\rm LR}}{d}= \frac{2-\eta_{\mathrm{SR}}}{d_{\mathrm{eff}}}\notag\\
    &\gamma_{\mathrm{LR}} = \gamma_{\mathrm{SR}},\quad\omega_{\mathrm{LR}}/d = \omega_{\mathrm{SR}}/d_{\mathrm{eff}}.\label{eq: critical exponents relations}
\end{align}
Interestingly, the critical exponents describing the scaling of global quantities, such as $\gamma$, shall directly correspond within the two theories, while the finite-size scaling exponents, such as $\nu$ only correspond once scaled via the effective dimension.

In order to obtain Eq.\,\eqref{eq: effective dim}, we assume that interactions do not shift the Gaussian estimate of the LR anomalous dimension  $\eta_{\mathrm{LR}} = 2-\sigma$. This assumption appears to be exact as it has been confirmed by perturbative arguments at $\mathcal{O}(\varepsilon^3)$\,\cite{Fisher1972critical}, functional RG studies\,\cite{Defenu2015fixedpoint,Balog2014critical}, Monte-Carlo\,\cite{Luijten2002boundary,Horita2017upper} and bootstrap calculations\,\cite{Behan2017scaling}. Combining this result with the relation between $\eta_{LR}$ and $\eta_{\mathrm{SR}}$ in Eq.\,\eqref{eq: critical exponents relations} leads to the dimensional identity in Eq.\,\eqref{eq: effective dim}. Based on the same argument one can derive the analytical expression for the threshold value $\sigma_{*}$, which coincides with the traditional result, first obtained by perturbative RG\,\cite{Sak1973recursion}, i.e. $\sigma_{*}=2-\eta_{\rm SR}$.

\section{Perturbative RG results}
\label{sec: Perturbative RG one loop results}
\begin{figure}
    \centering    \includegraphics[width=0.9\linewidth]{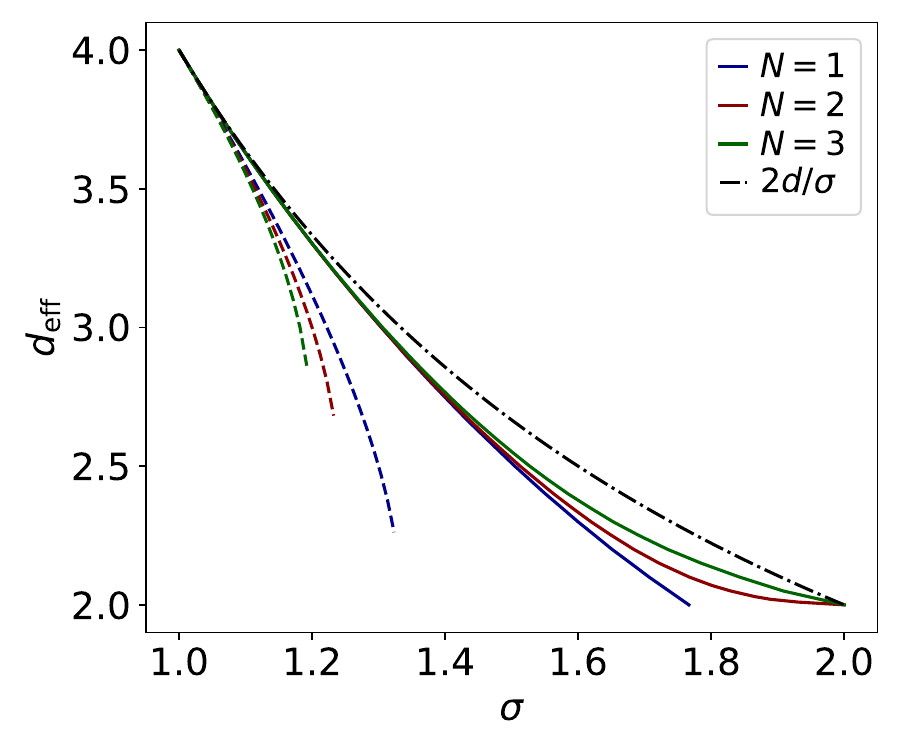}
    \caption{Effective dimension $d_{\mathrm{eff}}$ of the  two-dimensional LR $\mathrm{O}(N)$ models with $N=1$ (blue), $N = 2$ (red) and $N = 3$ (green), plotted as a function of the LR power law exponent $\sigma$. The results are computed using functional RG (solid lines), perturbative RG (dashed lines), and at the mean field level (dot dashed line).}
   \label{fig: deff_nu_LR_FRG}
\end{figure}

Using know perturbative RG results, one can obtain an explicit formula for $d_{\rm eff}$ valid up to corrections of order $\mathcal{O}(\varepsilon^3)$. Indeed, the epsilon expansion result \cite{Zinn2021quantum} for the $\eta$ exponent of the local model in dimension $d_{\mathrm{eff}}$ reads
\begin{align}
\label{per_eta_sr}
\eta_{\mathrm{SR}}^{d_{\mathrm{eff}}} = \frac{N + 2}{2(N + 8)}(4 - d_{\mathrm{eff}})^2 + \mathcal{O}((4 - d_{\mathrm{eff}})^3).
\end{align}
By inserting this expression into Eq.\,\eqref{eq: effective dim} one can solve for the effective dimension $d_{_\mathrm{eff}}$ and obtain two possible solutions. Then, in order to discriminate the physical solution of $d_{\rm eff}$, the correct mean-field threshold $\sigma = d/2$ has to be imposed at the upper critical dimension, i.e., $d_{\mathrm{eff}}(\sigma=d/2) = 4$. Doing so, the explicit expression is obtained as
\begin{align}
d_{\mathrm{eff}} \approx 4 - \frac{1}{d}\frac{N + 8}{N + 2}\left[\sigma - \sqrt{\sigma^2 + \frac{(4d^2 - 8d\sigma)(N + 2)}{(N + 8)}}\right],\label{eq: effective dimension perturbative}
\end{align}
which is valid for any $N$ and for values of $\sigma$ such that
\begin{align}
    d/2\leq\sigma\ll \frac{2d(4+2N-\sqrt{3N(N+2)})}{N+8}.\label{eq: region}
\end{align}
For larger values of $\sigma$, the perturbative expression looses its validity as Eq.\,\eqref{eq: effective dimension perturbative} becomes complex, indicating the need to go to higher orders in the $\varepsilon$-expansion.

Finally, by computing the values of the critical exponents of the local model in the $\varepsilon$-expansion up to $\mathcal{O}(\varepsilon^3)$ and using the relations with the LR exponents in Eq.\,\eqref{eq: critical exponents relations} and the expression for the effective dimension in Eq.\,\eqref{eq: effective dimension perturbative}, one immediately obtains the values of the LR exponents as a function of $\sigma$ and $d$, at the same level of approximation. 

We observe that the $\varepsilon$-expansion results results provide a good approximation only near $\sigma = d/2 = 1$,  as expected because in this region $d_\mathrm{eff}\approx 4$, see the dashed lines in Fig.\,\ref{fig: deff_nu_LR_FRG}. This corresponds to the region of validity of the $\mathcal{O}(\varepsilon^3)$ approximation determined by Eq.\,\eqref{eq: region}. Consequently, for these values of $\sigma$, the effective dimension approach provides a straightforward method to generalize $\varepsilon$-expansion results for the SR model to the LR case, as expected from the fact that the effective dimension equivalence is exact at the one-loop level.

Furthermore, we notice that for continuous theories with $N>1$, the mean field results $d_{\mathrm{eff}} = 2d/\sigma$ and $\nu_{LR} = d-\sigma$, which become exact in the $N\to\infty$ limit, perfectly interpolate between the exact results in $d = 2$ and $d = 4$, see the red and green solid lines in Fig.\,\ref{fig: deff_nu_LR_FRG}. On the other hand, the Ising case $N = 1$ (blue solid line in Fig.\,\ref{fig: deff_nu_LR_FRG}) shows a peculiar behavior and reaches $d_\mathrm{eff} = 2$ for $\sigma = \sigma^* = 7/4$. This is due to the fact that the Ising model, having a discrete symmetry, has a finite anomalous dimension even in $d = 2$. 

\section{Functional RG results}\label{sec: Functional RG results}
\label{sec_frg}
\begin{figure}
    \centering    \includegraphics[width=0.9\linewidth]{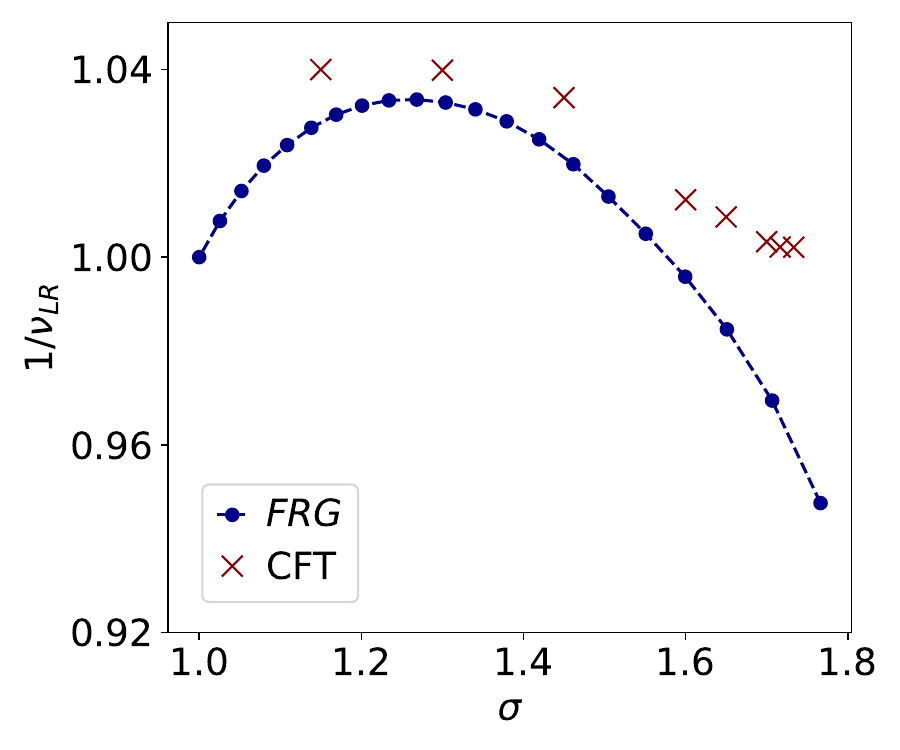}
    \caption{The correlation length exponent $\nu_{\rm LR}$ of the LR Ising model obtained via the functional RG approach within the LPA framework, see Eq.\,\eqref{eq: flow eq effective potential} is compared with accurate conformal bootstrap (CFT) result. The accuracy of functional RG in reproducing the CFT data lies within 92\%.}
   \label{fig: nu_FRG}
\end{figure}
Functional RG is a modern RG framework, which allows to derive an \emph{in principle} exact equation for the flow of the effective average action $\Gamma_k$ of the model under study. This framework, originated by the seminal work of Wilson and\,\cite{Wilson1983renormalization} Polchinski\,\cite{Polchinksi1984renormalization},  is more conveniently framed in terms of the Wetterich equation\,\cite{Wetterich1993exact}
\begin{align}
\label{wetterich_eq}
    \partial_t\Gamma_k = \frac{1}{2}\mathrm{Tr}\left[\frac{\partial_t R_k}{\Gamma^{(2)}+R_k}\right].
\end{align}
where $t =
\ln(k/k_0)$, $\Gamma^{(2)}$ is the second derivative of the effective action with respect to the order parameters, and $R_k(q)$ is a momentum space regulator function that cuts off the infrared divergences caused by slow modes $q\ll k$, while leaving the high-momentum modes $q\gg k$ almost untouched.

In order to treat Eq.\,\eqref{wetterich_eq} on has to project it on a restricted functional space, parametrised by a limited number of functional operators. In this perspective, a convenient ansatz for the effective action of the LR interacting $\mathrm{O}(N)$ model reads\,\cite{Defenu2015fixedpoint}
\begin{align}
\label{lr_ansatz}
    \Gamma_k[\varphi] = \int d^dx [Z_k\partial_\mu^{\frac{\sigma}{2}}\varphi_i\partial_\mu^{\frac{\sigma}{2}}\varphi_i+U(\rho)]
\end{align}
where  $\varphi_i$ is the $i$th component of $\boldsymbol{\varphi}$, $\rho = \varphi_i\varphi_i/2$ and the summation over repeated indexes is assumed. The notation $\partial_\mu^{\frac{\sigma}{2}}$ indicates that the inverse propagator of the effective action in Fourier space depends on $q^\sigma$. Then, introducing the dimensionless variables
\begin{align}
    \bar{U}_k(\bar{\rho}) = k^{-d}U_k(\rho),\quad\bar{\rho} = Z_kk^{d-\sigma}\rho,\quad \bar{q} = k^{-1}q,
\end{align}
and defining the generalized Litim cutoff \cite{Litim2002critical} suited for LR interactions \cite{Defenu2015fixedpoint,Defenu2016anisotropic,Defenu2017criticality}
\begin{align}
    R_k(q) = Z_k(k^\sigma-q^\sigma)\theta(k^\sigma-q^\sigma),
\end{align}
we obtain the following flow equation for the effective potential
\begin{align}
    \partial_t\bar{U}_k = &-d\bar{U}_k(\bar{\rho})+(d-\sigma+\delta\eta)\bar{\rho}\bar{U}_k'(\bar{\rho})\notag\\
    &+\frac{\sigma}{2}c_d(N-1)\frac{d+\sigma-\delta\eta}{(d+\sigma)(1+\bar{U}_k'(\bar{\rho}))}\notag\\
    &+\frac{\sigma}{2}c_d(N-1)\frac{d+\sigma-\delta\eta}{(d+\sigma)(1+\bar{U}_k'(\bar{\rho}+2\bar{\rho}\bar{U}_k''(\bar{\rho})))},\label{eq: flow eq effective potential}
\end{align}
where $c_d^{-1} = (4\pi)^{d/2}\Gamma(d/2 + 1)$ and $\delta\eta = -\partial_t \ln Z_k$.
Allowing the wavefunction renormalization $Z_k$ to be a running but field-independent coupling, we find its flow equation to be 
\begin{align}
    \partial_tZ_k = \lim_{p\to 0}\frac{d}{dp^\sigma}\partial_t\Gamma^{(2)}_k(p,-p). 
\end{align}
However, since the flow equation generates no non-analytic terms in $p$,  we find from its definition that $\delta\eta = 0$, in agreement with Sak’s picture\,\cite{Sak1973recursion}. Consequently, in Eq.\,\eqref{eq: flow eq effective potential}, we can omit the $\delta\eta$ terms.

Also in the functional RG framework, one can establish a mapping, between the LR critical exponents in $d$ dimensions and the equivalent local ones at the effective dimension $d_{\rm eff}$. For the correlation length critical exponent $\nu$, the correspondance is obtained by writing an eigenvalue equation for the stability of the perturbations around $\bar{U}_k = \bar{U}^*_k(\bar{\rho})$, which is the fixed point solution of Eq.\,\eqref{eq: flow eq effective potential}. Then making the substitution
\begin{align}
   \bar{U}_k(\bar{\rho}) = \bar{U}^*_k(\bar{\rho})+k^y \bar{u}_k(\bar{\rho}),  
\end{align}
one obtains a functional equation for the stability matrix exponents $y$s of the RG flow, which are related to the correlation length critical exponent via the relation $\nu^{-1} =\min\{y\}$.

Next, comparing Eq.\,\eqref{eq: flow eq effective potential}, together with its stability matrix extension, with their local counterparts in $d_{\mathrm{eff}}$ dimensions\,\cite{Codello2015critical},  and after reabsorbing the constant $c_d$ into the definition of the field\,\cite{morris1994truncations} one obtains again the dimensional equivalence in Eq.\eqref{eq: effective dim}, see Ref.\,\cite{Defenu2015fixedpoint} for the details of this derivation. Therefore, the dimensional correspondence described by the relations in Eq.\,\eqref{eq: critical exponents relations} can be recovered outside the euristic scaling theory framework of Sec.\,\ref{sec: derivation of deff}, suggesting it could be a highly accurate approximation for the real critical exponents of the LR model. 

The approximate nature of the effective dimension correspondence becomes evident going beyond the ansatz in Eq.\,\eqref{lr_ansatz} and including a local kinetic term
\begin{align}
\label{lr_ansatz_2}
    \Gamma_k[\varphi] = \int d^dx [\partial_\mu^{\frac{\sigma}{2}}\varphi_i\partial_\mu^{\frac{\sigma}{2}}\varphi_i+Z_{k}\partial_\mu\varphi_i\partial_\mu\varphi_i+U(\rho)]
\end{align}
where we omitted the running wave-function renormalization for the non-analytic kinetic term as we have already shown that it is irrelevant.

The study of the ansatz\,\eqref{lr_ansatz_2} has been performed in Ref.\,\cite{Defenu2015fixedpoint}, where it was shown that the resulting flow equation are not consistent with the effective dimension relations contrary to the simpler case of Eq.\,\eqref{eq: flow eq effective potential}. Nonetheless, the correction between the numerical values obtained within the two ansatz in Eq.\,\eqref{lr_ansatz} and Eq.\,\eqref{lr_ansatz_2} has been shown to be considerably small, remaining well below $5\%$ at all values of $\sigma$, reinforcing our expectation on the high accuracy of the effective dimension relation\,\cite{Defenu2015fixedpoint}.

A first confirmation of the above expectation follows from the comparison of the  numerical estimate for $\nu_{\rm LR}$ obtained through the study of Eq.\,\eqref{eq: flow eq effective potential} with the (possibly) exact results which have been recently obtained through conformal bootstrap\,\cite{Behan2024analytic}, see Fig.\,\ref{fig: nu_FRG}. The figure clearly shows that despite the rather crude approximation defined by Eq.\,\eqref{lr_ansatz} the functional RG results (full blue circles) essentially capture the trend of the CFT data, with the numerical error remaining confined within $7\%$. However, as we will argue in the next section, the effective dimension correspondence actually overcomes the functional RG accuracy when applied to the exact numerical estimates.

\section{Comparison with exact numerics}\label{sec: Comparison with exact numerics} 
\begin{figure}
    \centering    \includegraphics[width=0.9\linewidth]{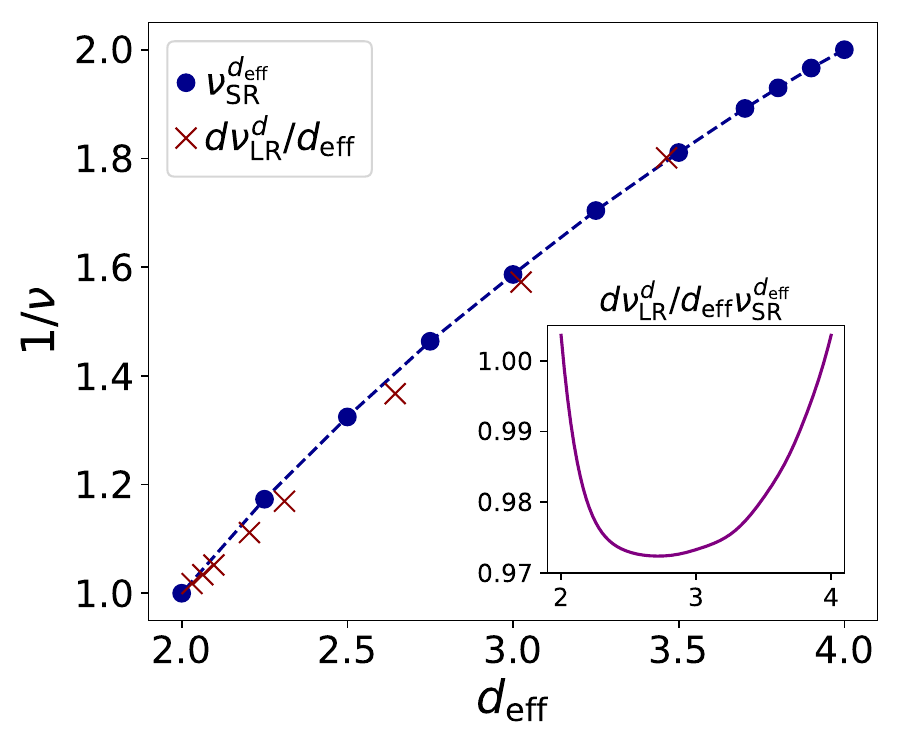}
    \caption{Comparison between the correlation length exponent of the local Ising model in $d_\mathrm{eff}$ dimension ($\nu_\mathrm{SR}^{d_\mathrm{eff}}$, blue dots) and that of the LR Ising model in $d=2$ dimensions, scaled by a factor $d/d_\mathrm{eff}$ ($d\nu_\mathrm{LR}^{d}/d_\mathrm{eff}$, red crosses), plotted as a function of the effective dimension $d_{\mathrm{eff}}(\sigma)$. The numerical data have been obtained through conformal bootstrap methods as reported in Refs.\,\cite{ElShowk2014conformal} and \,\cite{Behan2024analytic}, respectively. The inset shows the accuracy of the effective dimension prediction, estimated as the ratio $d\nu_{\mathrm{LR}}^d/d_{\mathrm{eff}}\nu_{\mathrm{SR}}^{d_{\mathrm{eff}}}$, plotted as a function of $d_{\mathrm{eff}}$.}
    \label{fig: effective_dimension_CFT}
\end{figure}
From the numerical side, the problem of validating the Sak scenario and the effective dimension correspondence has generated a long-standing debate. The first Monte Carlo (MC) studies of the problem, based on MC algorithms specific for LR interactions\,\cite{Luijten1995MonteCarlo}, supported $\sigma_{*}=2-\eta_{\rm SR}$\,\cite{Luijten2002boundary}. However, more recent MC results challenged Sak's scenario\,\cite{Blanchard2013influence}, reporting $\sigma_{*}=2$. Additionally, MC results for a percolation model with LR probabilities\,\cite{Grassberger2013TwoDimensional} basically mimic the picture of Ref.\,\cite{Blanchard2013influence} and do not reproduce Sak's result, although not explicitly discussing it. Furthermore, MC results for the Ising model with LR interaction in $d = 2$ presented in Ref.\cite{Angelini2014relations} evidenced the presence of logarithmic corrections in the correlation function when $\sigma$ is very close to the $\sigma_{*} = 2-\eta_{\mathrm{SR}}$ boundary. This implies numerical difficulty in extracting reliable results for the critical exponents with small error bars around $\sigma_{*} = 2-\eta_{\mathrm{SR}}$. While theoretical investigations now, almost unanimously, support Sak's picture\,\cite{Defenu2015fixedpoint,Behan2017scaling}, a proper estimate of the accuracy of the effective dimension correspondence remain open.

Here, we asses this accuracy by comparing numerically exact estimate for the correlation length exponent of the Ising model. These comparisons are, indeed, possible only for the Ising $N = 1$ case, where conformal bootstrap techniques have been recently used to obtain highly accurate estimates of the scaling dimensions in the LR case\,\cite{Behan2024analytic}. While the study of the nearest-neighbour Ising model in generic fractional dimension has been already available for almost a decade\,\cite{ElShowk2014conformal}. 

These data are compared in Fig. \ref{fig: effective_dimension_CFT}, where the exponent $1/\nu_{\mathrm{SR}}^{d_\mathrm{eff}}$ (blue dots) is plotted as a function of $d_\mathrm{eff}$ and compared with the corresponding LR exponent in two dimensions $1/\nu_{\mathrm{LR}}^{d}$ (red crosses) appropriately rescaled according to Eq.\,\eqref{eq: critical exponents relations}. While the match between the curves is not perfect, confirming the approximate nature of the effective dimension paradigm, the accuracy of the effective dimension approach clearly overcomes the prediction from the previous one-loop arguments. Indeed, as shown in the inset of Fig. \ref{fig: effective_dimension_CFT} the ratio $d\nu_{\mathrm{LR}}^d/d_{\mathrm{eff}}\nu_{\mathrm{SR}}^{d_{\mathrm{eff}}}$ is always larger then $\approx 0.97$, indicating a $97\%$ accuracy in the estimates for the LR exponents obtained using the effective dimension correspondence, well beyond the accuracy of the functional RG argument used to derive the effective dimension relation in Sec.\,\ref{sec_frg}.

It is worth noting that the Ising model ($N=1$) is expected to represent the worst-case scenario for the effective dimension prediction among the $\mathrm{O}(N)$ models. Indeed, the corrections to the effective dimension relation arise due to high-order momentum corrections to the vertexes of $\mathrm{O}(N)$ field theories, which are not parametrized by the ansatz\,\eqref{lr_ansatz}. These same vertex corrections cause the emergence of a finite anomalous dimension in local theories. Thus, the larger the anomalous dimension of a local theory the lower we estimate the accuracy of the effective dimension correspondence with its LR counterpart. As the Ising model displays the largest anomalous dimension within $\mathrm{O}(N)$ field theories we expect the effective dimension correspondence to be the most accurate as $N$ grows.

\section{Conclusion}
In this study, we have explored the critical behavior of LR interacting systems through a combination of perturbative and functional renormalization group (RG) approaches. Our primary focus was on establishing and validating the effective dimension framework, which relates the critical properties of LR models to those of local models in a suitably defined effective dimension $d_{\mathrm{eff}}$.

Firstly, we revisited Sak's seminal results using perturbative RG techniques. We confirmed that the effective dimension approach holds true at leading order in perturbation theory, providing a robust approximation for the critical exponents of LR models. This approximation is particularly accurate near the mean-field threshold $\sigma = d/2$ and remains valid up to $\mathcal{O}(\varepsilon^3)$ corrections. Our analysis showed that the effective dimension $d_{\mathrm{eff}}$ and the corresponding critical exponents can be explicitly computed, offering an efficient method for estimating the critical properties of LR systems.

Secondly, we extended our investigation to functional RG methods, which, allow for a more comprehensive treatment beyond the limitations of perturbative expansions. By deriving flow equations for the effective potential and the wavefunction renormalization, we demonstrated that the effective dimension approach is consistent with functional RG results. This consistency underscores the reliability of the effective dimension as a predictive tool for LR critical phenomena, even when interactions and fluctuations are considered more rigorously.

Finally, we compared the predictions of the effective dimension approach with exact numerical data obtained from conformal bootstrap methods for the two-dimensional Ising model with LR interactions. The comparison indicated excellent agreement, with an accuracy larger than $97\%$. 

In conclusion, our study provides strong evidence that the effective dimension approach, while approximate, offers an accurate and practical framework for estimating critical exponents in LR interacting models. It bridges the gap between perturbative RG predictions and exact numerical results, offering a comprehensive understanding of the universal features of critical phenomena in systems with LR interactions. 

\begin{acknowledgments}
 We thank Connor Behan for sharing the conformal bootstrap numerical data for the two-dimensional Ising model with LR interactions. N.D. acknowledges funding by the Swiss National Science Foundation (SNSF) under project funding ID: 200021 207537 and by the Deutsche Forschungsgemeinschaft (DFG, German Research Foundation) under Germany’s Excellence Strategy EXC2181/1-390900948 (the Heidelberg STRUCTURES Excellence Cluster).
\end{acknowledgments}

\begin{thebibliography}{33}%
\makeatletter
\providecommand \@ifxundefined [1]{%
 \@ifx{#1\undefined}
}%
\providecommand \@ifnum [1]{%
 \ifnum #1\expandafter \@firstoftwo
 \else \expandafter \@secondoftwo
 \fi
}%
\providecommand \@ifx [1]{%
 \ifx #1\expandafter \@firstoftwo
 \else \expandafter \@secondoftwo
 \fi
}%
\providecommand \natexlab [1]{#1}%
\providecommand \enquote  [1]{``#1''}%
\providecommand \bibnamefont  [1]{#1}%
\providecommand \bibfnamefont [1]{#1}%
\providecommand \citenamefont [1]{#1}%
\providecommand \href@noop [0]{\@secondoftwo}%
\providecommand \href [0]{\begingroup \@sanitize@url \@href}%
\providecommand \@href[1]{\@@startlink{#1}\@@href}%
\providecommand \@@href[1]{\endgroup#1\@@endlink}%
\providecommand \@sanitize@url [0]{\catcode `\\12\catcode `\$12\catcode `\&12\catcode `\#12\catcode `\^12\catcode `\_12\catcode `\%12\relax}%
\providecommand \@@startlink[1]{}%
\providecommand \@@endlink[0]{}%
\providecommand \url  [0]{\begingroup\@sanitize@url \@url }%
\providecommand \@url [1]{\endgroup\@href {#1}{\urlprefix }}%
\providecommand \urlprefix  [0]{URL }%
\providecommand \Eprint [0]{\href }%
\providecommand \doibase [0]{https://doi.org/}%
\providecommand \selectlanguage [0]{\@gobble}%
\providecommand \bibinfo  [0]{\@secondoftwo}%
\providecommand \bibfield  [0]{\@secondoftwo}%
\providecommand \translation [1]{[#1]}%
\providecommand \BibitemOpen [0]{}%
\providecommand \bibitemStop [0]{}%
\providecommand \bibitemNoStop [0]{.\EOS\space}%
\providecommand \EOS [0]{\spacefactor3000\relax}%
\providecommand \BibitemShut  [1]{\csname bibitem#1\endcsname}%
\let\auto@bib@innerbib\@empty
\bibitem [{\citenamefont {Wilson}(1971{\natexlab{a}})}]{Wilson1971renormalizationI}%
  \BibitemOpen
  \bibfield  {author} {\bibinfo {author} {\bibfnamefont {K.~G.}\ \bibnamefont {Wilson}},\ }\bibfield  {title} {\bibinfo {title} {Renormalization group and critical phenomena. i. renormalization group and the kadanoff scaling picture},\ }\href {https://doi.org/10.1103/PhysRevB.4.3174} {\bibfield  {journal} {\bibinfo  {journal} {Phys. Rev. B}\ }\textbf {\bibinfo {volume} {4}},\ \bibinfo {pages} {3174} (\bibinfo {year} {1971}{\natexlab{a}})}\BibitemShut {NoStop}%
\bibitem [{\citenamefont {Wilson}(1971{\natexlab{b}})}]{Wilson1971renormalizationII}%
  \BibitemOpen
  \bibfield  {author} {\bibinfo {author} {\bibfnamefont {K.~G.}\ \bibnamefont {Wilson}},\ }\bibfield  {title} {\bibinfo {title} {Renormalization group and critical phenomena. ii. phase-space cell analysis of critical behavior},\ }\href {https://doi.org/10.1103/PhysRevB.4.3184} {\bibfield  {journal} {\bibinfo  {journal} {Phys. Rev. B}\ }\textbf {\bibinfo {volume} {4}},\ \bibinfo {pages} {3184} (\bibinfo {year} {1971}{\natexlab{b}})}\BibitemShut {NoStop}%
\bibitem [{\citenamefont {Zinn-Justin}(2021)}]{Zinn2021quantum}%
  \BibitemOpen
  \bibfield  {author} {\bibinfo {author} {\bibfnamefont {J.}~\bibnamefont {Zinn-Justin}},\ }\href {https://books.google.it/books?id=ioskEAAAQBAJ} {\emph {\bibinfo {title} {Quantum Field Theory and Critical Phenomena}}},\ International series of monographs on physics\ (\bibinfo  {publisher} {Oxford University Press},\ \bibinfo {year} {2021})\BibitemShut {NoStop}%
\bibitem [{\citenamefont {Pelissetto}\ and\ \citenamefont {Vicari}(2002)}]{Pellissetto2002critical}%
  \BibitemOpen
  \bibfield  {author} {\bibinfo {author} {\bibfnamefont {A.}~\bibnamefont {Pelissetto}}\ and\ \bibinfo {author} {\bibfnamefont {E.}~\bibnamefont {Vicari}},\ }\bibfield  {title} {\bibinfo {title} {Critical phenomena and renormalization-group theory},\ }\href {https://doi.org/https://doi.org/10.1016/S0370-1573(02)00219-3} {\bibfield  {journal} {\bibinfo  {journal} {Physics Reports}\ }\textbf {\bibinfo {volume} {368}},\ \bibinfo {pages} {549} (\bibinfo {year} {2002})}\BibitemShut {NoStop}%
\bibitem [{\citenamefont {Campa}\ \emph {et~al.}(2009)\citenamefont {Campa}, \citenamefont {Dauxois},\ and\ \citenamefont {Ruffo}}]{Campa2009Statistical}%
  \BibitemOpen
  \bibfield  {author} {\bibinfo {author} {\bibfnamefont {A.}~\bibnamefont {Campa}}, \bibinfo {author} {\bibfnamefont {T.}~\bibnamefont {Dauxois}},\ and\ \bibinfo {author} {\bibfnamefont {S.}~\bibnamefont {Ruffo}},\ }\bibfield  {title} {\bibinfo {title} {Statistical mechanics and dynamics of solvable models with long-range interactions},\ }\href {https://doi.org/https://doi.org/10.1016/j.physrep.2009.07.001} {\bibfield  {journal} {\bibinfo  {journal} {Physics Reports}\ }\textbf {\bibinfo {volume} {480}},\ \bibinfo {pages} {57} (\bibinfo {year} {2009})}\BibitemShut {NoStop}%
\bibitem [{\citenamefont {Campa}\ \emph {et~al.}(2014)\citenamefont {Campa}, \citenamefont {Dauxois}, \citenamefont {Fanelli},\ and\ \citenamefont {Ruffo}}]{Campa2014physics}%
  \BibitemOpen
  \bibfield  {author} {\bibinfo {author} {\bibfnamefont {A.}~\bibnamefont {Campa}}, \bibinfo {author} {\bibfnamefont {T.}~\bibnamefont {Dauxois}}, \bibinfo {author} {\bibfnamefont {D.}~\bibnamefont {Fanelli}},\ and\ \bibinfo {author} {\bibfnamefont {S.}~\bibnamefont {Ruffo}},\ }\href {https://books.google.it/books?id=AIw_BAAAQBAJ} {\emph {\bibinfo {title} {Physics of Long-range Interacting Systems}}}\ (\bibinfo  {publisher} {Oxford University Press},\ \bibinfo {year} {2014})\BibitemShut {NoStop}%
\bibitem [{\citenamefont {Defenu}\ \emph {et~al.}(2023)\citenamefont {Defenu}, \citenamefont {Donner}, \citenamefont {Macr\`{\i}}, \citenamefont {Pagano}, \citenamefont {Ruffo},\ and\ \citenamefont {Trombettoni}}]{Defenu2023Longrange}%
  \BibitemOpen
  \bibfield  {author} {\bibinfo {author} {\bibfnamefont {N.}~\bibnamefont {Defenu}}, \bibinfo {author} {\bibfnamefont {T.}~\bibnamefont {Donner}}, \bibinfo {author} {\bibfnamefont {T.}~\bibnamefont {Macr\`{\i}}}, \bibinfo {author} {\bibfnamefont {G.}~\bibnamefont {Pagano}}, \bibinfo {author} {\bibfnamefont {S.}~\bibnamefont {Ruffo}},\ and\ \bibinfo {author} {\bibfnamefont {A.}~\bibnamefont {Trombettoni}},\ }\bibfield  {title} {\bibinfo {title} {Long-range interacting quantum systems},\ }\href {https://doi.org/10.1103/RevModPhys.95.035002} {\bibfield  {journal} {\bibinfo  {journal} {Rev. Mod. Phys.}\ }\textbf {\bibinfo {volume} {95}},\ \bibinfo {pages} {035002} (\bibinfo {year} {2023})}\BibitemShut {NoStop}%
\bibitem [{\citenamefont {Mermin}\ and\ \citenamefont {Wagner}(1966)}]{Mermin1966absence}%
  \BibitemOpen
  \bibfield  {author} {\bibinfo {author} {\bibfnamefont {N.~D.}\ \bibnamefont {Mermin}}\ and\ \bibinfo {author} {\bibfnamefont {H.}~\bibnamefont {Wagner}},\ }\bibfield  {title} {\bibinfo {title} {Absence of ferromagnetism or antiferromagnetism in one- or two-dimensional isotropic heisenberg models},\ }\href {https://doi.org/10.1103/PhysRevLett.17.1133} {\bibfield  {journal} {\bibinfo  {journal} {Phys. Rev. Lett.}\ }\textbf {\bibinfo {volume} {17}},\ \bibinfo {pages} {1133} (\bibinfo {year} {1966})}\BibitemShut {NoStop}%
\bibitem [{\citenamefont {Fisher}\ \emph {et~al.}(1972)\citenamefont {Fisher}, \citenamefont {Ma},\ and\ \citenamefont {Nickel}}]{Fisher1972critical}%
  \BibitemOpen
  \bibfield  {author} {\bibinfo {author} {\bibfnamefont {M.~E.}\ \bibnamefont {Fisher}}, \bibinfo {author} {\bibfnamefont {S.-k.}\ \bibnamefont {Ma}},\ and\ \bibinfo {author} {\bibfnamefont {B.~G.}\ \bibnamefont {Nickel}},\ }\bibfield  {title} {\bibinfo {title} {Critical exponents for long-range interactions},\ }\href {https://doi.org/10.1103/PhysRevLett.29.917} {\bibfield  {journal} {\bibinfo  {journal} {Phys. Rev. Lett.}\ }\textbf {\bibinfo {volume} {29}},\ \bibinfo {pages} {917} (\bibinfo {year} {1972})}\BibitemShut {NoStop}%
\bibitem [{\citenamefont {Defenu}\ \emph {et~al.}(2020)\citenamefont {Defenu}, \citenamefont {Codello}, \citenamefont {Ruffo},\ and\ \citenamefont {Trombettoni}}]{Defenu2020criticality}%
  \BibitemOpen
  \bibfield  {author} {\bibinfo {author} {\bibfnamefont {N.}~\bibnamefont {Defenu}}, \bibinfo {author} {\bibfnamefont {A.}~\bibnamefont {Codello}}, \bibinfo {author} {\bibfnamefont {S.}~\bibnamefont {Ruffo}},\ and\ \bibinfo {author} {\bibfnamefont {A.}~\bibnamefont {Trombettoni}},\ }\bibfield  {title} {\bibinfo {title} {Criticality of spin systems with weak long-range interactions},\ }\href {https://doi.org/10.1088/1751-8121/ab6a6c} {\bibfield  {journal} {\bibinfo  {journal} {Journal of Physics A: Mathematical and Theoretical}\ }\textbf {\bibinfo {volume} {53}},\ \bibinfo {pages} {143001} (\bibinfo {year} {2020})}\BibitemShut {NoStop}%
\bibitem [{\citenamefont {Kotliar}\ \emph {et~al.}(1983)\citenamefont {Kotliar}, \citenamefont {Anderson},\ and\ \citenamefont {Stein}}]{Kotliar1983one}%
  \BibitemOpen
  \bibfield  {author} {\bibinfo {author} {\bibfnamefont {G.}~\bibnamefont {Kotliar}}, \bibinfo {author} {\bibfnamefont {P.~W.}\ \bibnamefont {Anderson}},\ and\ \bibinfo {author} {\bibfnamefont {D.~L.}\ \bibnamefont {Stein}},\ }\bibfield  {title} {\bibinfo {title} {One-dimensional spin-glass model with long-range random interactions},\ }\href {https://doi.org/10.1103/PhysRevB.27.602} {\bibfield  {journal} {\bibinfo  {journal} {Phys. Rev. B}\ }\textbf {\bibinfo {volume} {27}},\ \bibinfo {pages} {602} (\bibinfo {year} {1983})}\BibitemShut {NoStop}%
\bibitem [{\citenamefont {Ba\~nos}\ \emph {et~al.}(2012)\citenamefont {Ba\~nos}, \citenamefont {Fernandez}, \citenamefont {Martin-Mayor},\ and\ \citenamefont {Young}}]{Banos2012correspondence}%
  \BibitemOpen
  \bibfield  {author} {\bibinfo {author} {\bibfnamefont {R.~A.}\ \bibnamefont {Ba\~nos}}, \bibinfo {author} {\bibfnamefont {L.~A.}\ \bibnamefont {Fernandez}}, \bibinfo {author} {\bibfnamefont {V.}~\bibnamefont {Martin-Mayor}},\ and\ \bibinfo {author} {\bibfnamefont {A.~P.}\ \bibnamefont {Young}},\ }\bibfield  {title} {\bibinfo {title} {Correspondence between long-range and short-range spin glasses},\ }\href {https://doi.org/10.1103/PhysRevB.86.134416} {\bibfield  {journal} {\bibinfo  {journal} {Phys. Rev. B}\ }\textbf {\bibinfo {volume} {86}},\ \bibinfo {pages} {134416} (\bibinfo {year} {2012})}\BibitemShut {NoStop}%
\bibitem [{\citenamefont {Angelini}\ \emph {et~al.}(2014)\citenamefont {Angelini}, \citenamefont {Parisi},\ and\ \citenamefont {Ricci-Tersenghi}}]{Angelini2014relations}%
  \BibitemOpen
  \bibfield  {author} {\bibinfo {author} {\bibfnamefont {M.~C.}\ \bibnamefont {Angelini}}, \bibinfo {author} {\bibfnamefont {G.}~\bibnamefont {Parisi}},\ and\ \bibinfo {author} {\bibfnamefont {F.}~\bibnamefont {Ricci-Tersenghi}},\ }\bibfield  {title} {\bibinfo {title} {Relations between short-range and long-range ising models},\ }\href {https://doi.org/10.1103/PhysRevE.89.062120} {\bibfield  {journal} {\bibinfo  {journal} {Phys. Rev. E}\ }\textbf {\bibinfo {volume} {89}},\ \bibinfo {pages} {062120} (\bibinfo {year} {2014})}\BibitemShut {NoStop}%
\bibitem [{\citenamefont {Joyce}(1966)}]{Joyce1966spherical}%
  \BibitemOpen
  \bibfield  {author} {\bibinfo {author} {\bibfnamefont {G.~S.}\ \bibnamefont {Joyce}},\ }\bibfield  {title} {\bibinfo {title} {Spherical model with long-range ferromagnetic interactions},\ }\href {https://doi.org/10.1103/PhysRev.146.349} {\bibfield  {journal} {\bibinfo  {journal} {Phys. Rev.}\ }\textbf {\bibinfo {volume} {146}},\ \bibinfo {pages} {349} (\bibinfo {year} {1966})}\BibitemShut {NoStop}%
\bibitem [{\citenamefont {Defenu}\ \emph {et~al.}(2015)\citenamefont {Defenu}, \citenamefont {Trombettoni},\ and\ \citenamefont {Codello}}]{Defenu2015fixedpoint}%
  \BibitemOpen
  \bibfield  {author} {\bibinfo {author} {\bibfnamefont {N.}~\bibnamefont {Defenu}}, \bibinfo {author} {\bibfnamefont {A.}~\bibnamefont {Trombettoni}},\ and\ \bibinfo {author} {\bibfnamefont {A.}~\bibnamefont {Codello}},\ }\bibfield  {title} {\bibinfo {title} {Fixed-point structure and effective fractional dimensionality for o($n$) models with long-range interactions},\ }\href {https://doi.org/10.1103/PhysRevE.92.052113} {\bibfield  {journal} {\bibinfo  {journal} {Phys. Rev. E}\ }\textbf {\bibinfo {volume} {92}},\ \bibinfo {pages} {052113} (\bibinfo {year} {2015})}\BibitemShut {NoStop}%
\bibitem [{\citenamefont {Behan}\ \emph {et~al.}(2017)\citenamefont {Behan}, \citenamefont {Rastelli}, \citenamefont {Rychkov},\ and\ \citenamefont {Zan}}]{Behan2017scaling}%
  \BibitemOpen
  \bibfield  {author} {\bibinfo {author} {\bibfnamefont {C.}~\bibnamefont {Behan}}, \bibinfo {author} {\bibfnamefont {L.}~\bibnamefont {Rastelli}}, \bibinfo {author} {\bibfnamefont {S.}~\bibnamefont {Rychkov}},\ and\ \bibinfo {author} {\bibfnamefont {B.}~\bibnamefont {Zan}},\ }\bibfield  {title} {\bibinfo {title} {A scaling theory for the long-range to short-range crossover and an infrared duality},\ }\href {https://doi.org/10.1088/1751-8121/aa8099} {\bibfield  {journal} {\bibinfo  {journal} {Journal of Physics A: Mathematical and Theoretical}\ }\textbf {\bibinfo {volume} {50}},\ \bibinfo {pages} {354002} (\bibinfo {year} {2017})}\BibitemShut {NoStop}%
\bibitem [{\citenamefont {Sak}(1973)}]{Sak1973recursion}%
  \BibitemOpen
  \bibfield  {author} {\bibinfo {author} {\bibfnamefont {J.}~\bibnamefont {Sak}},\ }\bibfield  {title} {\bibinfo {title} {Recursion relations and fixed points for ferromagnets with long-range interactions},\ }\href {https://doi.org/10.1103/PhysRevB.8.281} {\bibfield  {journal} {\bibinfo  {journal} {Phys. Rev. B}\ }\textbf {\bibinfo {volume} {8}},\ \bibinfo {pages} {281} (\bibinfo {year} {1973})}\BibitemShut {NoStop}%
\bibitem [{\citenamefont {Balog}\ \emph {et~al.}(2014)\citenamefont {Balog}, \citenamefont {Tarjus},\ and\ \citenamefont {Tissier}}]{Balog2014critical}%
  \BibitemOpen
  \bibfield  {author} {\bibinfo {author} {\bibfnamefont {I.}~\bibnamefont {Balog}}, \bibinfo {author} {\bibfnamefont {G.}~\bibnamefont {Tarjus}},\ and\ \bibinfo {author} {\bibfnamefont {M.}~\bibnamefont {Tissier}},\ }\bibfield  {title} {\bibinfo {title} {Critical behaviour of the random-field ising model with long-range interactions in one dimension},\ }\href {https://doi.org/10.1088/1742-5468/2014/10/p10017} {\bibfield  {journal} {\bibinfo  {journal} {Journal of Statistical Mechanics: Theory and Experiment}\ }\textbf {\bibinfo {volume} {2014}},\ \bibinfo {pages} {P10017} (\bibinfo {year} {2014})}\BibitemShut {NoStop}%
\bibitem [{\citenamefont {Luijten}\ and\ \citenamefont {Bl\"ote}(2002)}]{Luijten2002boundary}%
  \BibitemOpen
  \bibfield  {author} {\bibinfo {author} {\bibfnamefont {E.}~\bibnamefont {Luijten}}\ and\ \bibinfo {author} {\bibfnamefont {H.~W.~J.}\ \bibnamefont {Bl\"ote}},\ }\bibfield  {title} {\bibinfo {title} {Boundary between long-range and short-range critical behavior in systems with algebraic interactions},\ }\href {https://doi.org/10.1103/PhysRevLett.89.025703} {\bibfield  {journal} {\bibinfo  {journal} {Phys. Rev. Lett.}\ }\textbf {\bibinfo {volume} {89}},\ \bibinfo {pages} {025703} (\bibinfo {year} {2002})}\BibitemShut {NoStop}%
\bibitem [{\citenamefont {Horita}\ \emph {et~al.}(2017)\citenamefont {Horita}, \citenamefont {Suwa},\ and\ \citenamefont {Todo}}]{Horita2017upper}%
  \BibitemOpen
  \bibfield  {author} {\bibinfo {author} {\bibfnamefont {T.}~\bibnamefont {Horita}}, \bibinfo {author} {\bibfnamefont {H.}~\bibnamefont {Suwa}},\ and\ \bibinfo {author} {\bibfnamefont {S.}~\bibnamefont {Todo}},\ }\bibfield  {title} {\bibinfo {title} {Upper and lower critical decay exponents of ising ferromagnets with long-range interaction},\ }\href {https://doi.org/10.1103/PhysRevE.95.012143} {\bibfield  {journal} {\bibinfo  {journal} {Phys. Rev. E}\ }\textbf {\bibinfo {volume} {95}},\ \bibinfo {pages} {012143} (\bibinfo {year} {2017})}\BibitemShut {NoStop}%
\bibitem [{\citenamefont {Wilson}(1983)}]{Wilson1983renormalization}%
  \BibitemOpen
  \bibfield  {author} {\bibinfo {author} {\bibfnamefont {K.~G.}\ \bibnamefont {Wilson}},\ }\bibfield  {title} {\bibinfo {title} {The renormalization group and critical phenomena},\ }\href {https://doi.org/10.1103/RevModPhys.55.583} {\bibfield  {journal} {\bibinfo  {journal} {Rev. Mod. Phys.}\ }\textbf {\bibinfo {volume} {55}},\ \bibinfo {pages} {583} (\bibinfo {year} {1983})}\BibitemShut {NoStop}%
\bibitem [{\citenamefont {Polchinski}(1984)}]{Polchinksi1984renormalization}%
  \BibitemOpen
  \bibfield  {author} {\bibinfo {author} {\bibfnamefont {J.}~\bibnamefont {Polchinski}},\ }\bibfield  {title} {\bibinfo {title} {Renormalization and effective lagrangians},\ }\href {https://doi.org/https://doi.org/10.1016/0550-3213(84)90287-6} {\bibfield  {journal} {\bibinfo  {journal} {Nuclear Physics B}\ }\textbf {\bibinfo {volume} {231}},\ \bibinfo {pages} {269} (\bibinfo {year} {1984})}\BibitemShut {NoStop}%
\bibitem [{\citenamefont {Wetterich}(1993)}]{Wetterich1993exact}%
  \BibitemOpen
  \bibfield  {author} {\bibinfo {author} {\bibfnamefont {C.}~\bibnamefont {Wetterich}},\ }\bibfield  {title} {\bibinfo {title} {Exact evolution equation for the effective potential},\ }\href {https://doi.org/https://doi.org/10.1016/0370-2693(93)90726-X} {\bibfield  {journal} {\bibinfo  {journal} {Physics Letters B}\ }\textbf {\bibinfo {volume} {301}},\ \bibinfo {pages} {90} (\bibinfo {year} {1993})}\BibitemShut {NoStop}%
\bibitem [{\citenamefont {Litim}(2002)}]{Litim2002critical}%
  \BibitemOpen
  \bibfield  {author} {\bibinfo {author} {\bibfnamefont {D.~F.}\ \bibnamefont {Litim}},\ }\bibfield  {title} {\bibinfo {title} {Critical exponents from optimised renormalisation group flows},\ }\href {https://doi.org/https://doi.org/10.1016/S0550-3213(02)00186-4} {\bibfield  {journal} {\bibinfo  {journal} {Nuclear Physics B}\ }\textbf {\bibinfo {volume} {631}},\ \bibinfo {pages} {128} (\bibinfo {year} {2002})}\BibitemShut {NoStop}%
\bibitem [{\citenamefont {Defenu}\ \emph {et~al.}(2016)\citenamefont {Defenu}, \citenamefont {Trombettoni},\ and\ \citenamefont {Ruffo}}]{Defenu2016anisotropic}%
  \BibitemOpen
  \bibfield  {author} {\bibinfo {author} {\bibfnamefont {N.}~\bibnamefont {Defenu}}, \bibinfo {author} {\bibfnamefont {A.}~\bibnamefont {Trombettoni}},\ and\ \bibinfo {author} {\bibfnamefont {S.}~\bibnamefont {Ruffo}},\ }\bibfield  {title} {\bibinfo {title} {Anisotropic long-range spin systems},\ }\href {https://doi.org/10.1103/PhysRevB.94.224411} {\bibfield  {journal} {\bibinfo  {journal} {Phys. Rev. B}\ }\textbf {\bibinfo {volume} {94}},\ \bibinfo {pages} {224411} (\bibinfo {year} {2016})}\BibitemShut {NoStop}%
\bibitem [{\citenamefont {Defenu}\ \emph {et~al.}(2017)\citenamefont {Defenu}, \citenamefont {Trombettoni},\ and\ \citenamefont {Ruffo}}]{Defenu2017criticality}%
  \BibitemOpen
  \bibfield  {author} {\bibinfo {author} {\bibfnamefont {N.}~\bibnamefont {Defenu}}, \bibinfo {author} {\bibfnamefont {A.}~\bibnamefont {Trombettoni}},\ and\ \bibinfo {author} {\bibfnamefont {S.}~\bibnamefont {Ruffo}},\ }\bibfield  {title} {\bibinfo {title} {Criticality and phase diagram of quantum long-range o($n$) models},\ }\href {https://doi.org/10.1103/PhysRevB.96.104432} {\bibfield  {journal} {\bibinfo  {journal} {Phys. Rev. B}\ }\textbf {\bibinfo {volume} {96}},\ \bibinfo {pages} {104432} (\bibinfo {year} {2017})}\BibitemShut {NoStop}%
\bibitem [{\citenamefont {Codello}\ \emph {et~al.}(2015)\citenamefont {Codello}, \citenamefont {Defenu},\ and\ \citenamefont {D'Odorico}}]{Codello2015critical}%
  \BibitemOpen
  \bibfield  {author} {\bibinfo {author} {\bibfnamefont {A.}~\bibnamefont {Codello}}, \bibinfo {author} {\bibfnamefont {N.}~\bibnamefont {Defenu}},\ and\ \bibinfo {author} {\bibfnamefont {G.}~\bibnamefont {D'Odorico}},\ }\bibfield  {title} {\bibinfo {title} {Critical exponents of $o(n)$ models in fractional dimensions},\ }\href {https://doi.org/10.1103/PhysRevD.91.105003} {\bibfield  {journal} {\bibinfo  {journal} {Phys. Rev. D}\ }\textbf {\bibinfo {volume} {91}},\ \bibinfo {pages} {105003} (\bibinfo {year} {2015})}\BibitemShut {NoStop}%
\bibitem [{\citenamefont {Morris}(1994)}]{morris1994truncations}%
  \BibitemOpen
  \bibfield  {author} {\bibinfo {author} {\bibfnamefont {T.~R.}\ \bibnamefont {Morris}},\ }\bibfield  {title} {\bibinfo {title} {On truncations of the exact renormalization group},\ }\href@noop {} {\bibfield  {journal} {\bibinfo  {journal} {Physics Letters B}\ }\textbf {\bibinfo {volume} {334}},\ \bibinfo {pages} {355} (\bibinfo {year} {1994})}\BibitemShut {NoStop}%
\bibitem [{\citenamefont {Behan}\ \emph {et~al.}(2024)\citenamefont {Behan}, \citenamefont {Lauria}, \citenamefont {Nocchi},\ and\ \citenamefont {van Vliet}}]{Behan2024analytic}%
  \BibitemOpen
  \bibfield  {author} {\bibinfo {author} {\bibfnamefont {C.}~\bibnamefont {Behan}}, \bibinfo {author} {\bibfnamefont {E.}~\bibnamefont {Lauria}}, \bibinfo {author} {\bibfnamefont {M.}~\bibnamefont {Nocchi}},\ and\ \bibinfo {author} {\bibfnamefont {P.}~\bibnamefont {van Vliet}},\ }\bibfield  {title} {\bibinfo {title} {Analytic and numerical bootstrap for the long-range ising model},\ }\href {https://doi.org/10.1007/JHEP03(2024)136} {\bibfield  {journal} {\bibinfo  {journal} {Journal of High Energy Physics}\ }\textbf {\bibinfo {volume} {2024}},\ \bibinfo {pages} {136} (\bibinfo {year} {2024})}\BibitemShut {NoStop}%
\bibitem [{\citenamefont {El-Showk}\ \emph {et~al.}(2014)\citenamefont {El-Showk}, \citenamefont {Paulos}, \citenamefont {Poland}, \citenamefont {Rychkov}, \citenamefont {Simmons-Duffin},\ and\ \citenamefont {Vichi}}]{ElShowk2014conformal}%
  \BibitemOpen
  \bibfield  {author} {\bibinfo {author} {\bibfnamefont {S.}~\bibnamefont {El-Showk}}, \bibinfo {author} {\bibfnamefont {M.}~\bibnamefont {Paulos}}, \bibinfo {author} {\bibfnamefont {D.}~\bibnamefont {Poland}}, \bibinfo {author} {\bibfnamefont {S.}~\bibnamefont {Rychkov}}, \bibinfo {author} {\bibfnamefont {D.}~\bibnamefont {Simmons-Duffin}},\ and\ \bibinfo {author} {\bibfnamefont {A.}~\bibnamefont {Vichi}},\ }\bibfield  {title} {\bibinfo {title} {Conformal field theories in fractional dimensions},\ }\href {https://doi.org/10.1103/PhysRevLett.112.141601} {\bibfield  {journal} {\bibinfo  {journal} {Phys. Rev. Lett.}\ }\textbf {\bibinfo {volume} {112}},\ \bibinfo {pages} {141601} (\bibinfo {year} {2014})}\BibitemShut {NoStop}%
\bibitem [{\citenamefont {Luijten}\ and\ \citenamefont {Bl\"{o}te}(1995)}]{Luijten1995MonteCarlo}%
  \BibitemOpen
  \bibfield  {author} {\bibinfo {author} {\bibfnamefont {E.}~\bibnamefont {Luijten}}\ and\ \bibinfo {author} {\bibfnamefont {H.~W.}\ \bibnamefont {Bl\"{o}te}},\ }\bibfield  {title} {\bibinfo {title} {Monte carlo method for spin models with long-range interactions},\ }\href {https://doi.org/10.1142/S0129183195000265} {\bibfield  {journal} {\bibinfo  {journal} {International Journal of Modern Physics C}\ }\textbf {\bibinfo {volume} {06}},\ \bibinfo {pages} {359} (\bibinfo {year} {1995})},\ \Eprint {https://arxiv.org/abs/https://doi.org/10.1142/S0129183195000265} {https://doi.org/10.1142/S0129183195000265} \BibitemShut {NoStop}%
\bibitem [{\citenamefont {Blanchard}\ \emph {et~al.}(2013)\citenamefont {Blanchard}, \citenamefont {Picco},\ and\ \citenamefont {Rajabpour}}]{Blanchard2013influence}%
  \BibitemOpen
  \bibfield  {author} {\bibinfo {author} {\bibfnamefont {T.}~\bibnamefont {Blanchard}}, \bibinfo {author} {\bibfnamefont {M.}~\bibnamefont {Picco}},\ and\ \bibinfo {author} {\bibfnamefont {M.~A.}\ \bibnamefont {Rajabpour}},\ }\bibfield  {title} {\bibinfo {title} {Influence of long-range interactions on the critical behavior of the ising model},\ }\href {https://doi.org/10.1209/0295-5075/101/56003} {\bibfield  {journal} {\bibinfo  {journal} {Europhysics Letters}\ }\textbf {\bibinfo {volume} {101}},\ \bibinfo {pages} {56003} (\bibinfo {year} {2013})}\BibitemShut {NoStop}%
\bibitem [{\citenamefont {Grassberger}(2013)}]{Grassberger2013TwoDimensional}%
  \BibitemOpen
  \bibfield  {author} {\bibinfo {author} {\bibfnamefont {P.}~\bibnamefont {Grassberger}},\ }\bibfield  {title} {\bibinfo {title} {Two-dimensional sir epidemics with long range infection},\ }\href {https://doi.org/10.1007/s10955-013-0824-7} {\bibfield  {journal} {\bibinfo  {journal} {Journal of Statistical Physics}\ }\textbf {\bibinfo {volume} {153}},\ \bibinfo {pages} {289} (\bibinfo {year} {2013})}\BibitemShut {NoStop}%
\end{thebibliography}
%

\end{document}